\documentclass[%
 aip,
 jmp,%
 amsmath,amssymb,
 reprint,%
]{revtex4-1}
\usepackage{dcolumn}
\usepackage{bm}
\usepackage[T1]{fontenc}
\usepackage[latin1]{inputenc}
\usepackage{epsfig}
\usepackage{amsmath,amssymb}
\usepackage{graphicx}
\usepackage{caption}
\usepackage{subcaption}

\def\XXint#1#2#3{{\setbox0=\hbox{$#1{#2#3}{\int}$}
    \vcenter{\hbox{$#2#3$}}\kern-.5\wd0}}

\def\11{\mathbf 1}
\begin{document}
\numberwithin{equation}{section}
\newtheorem{theoreme}{Theorem}[section]
\newtheorem{proposition}[theoreme]{Proposition}
\newtheorem{remarque}[theoreme]{Remark}
\newtheorem{lemme}[theoreme]{Lemma}
\newtheorem{corollaire}[theoreme]{Corollary}
\newtheorem{definition}[theoreme]{Definition}

\title[Ion kinetic effects on the ignition and burn in ICF]{Ion kinetic effects on the ignition and burn of ICF targets}
\author{B.E Peigney}
\affiliation{CEA/DIF, 91297 Arpajon Cedex, France}
\author{O. Larroche}
\affiliation{CEA/DIF, 91297 Arpajon Cedex, France}
\author{V. Tikhonchuk}
\affiliation{University Bordeaux-CNRS-CEA-CELIA, 33405 Talence Cedex, France}

\begin{abstract}
In this Article,  we study the hydrodynamics and burn of the thermonuclear fuel in inertial confinement
fusion pellets at the ion kinetic level. The analysis is based on a two-velocity-scale
Vlasov-Fokker-Planck kinetic model that is specially tailored to treat fusion products (suprathermal $\alpha$-particles) in a self-consistent manner with the thermal bulk. The model assumes  spherical symmetry in configuration space and axial symmetry in velocity space around the mean flow velocity.  Compared to fluid simulations where a multi-group diffusion scheme is applied to model $\alpha$ transport, the full ion-kinetic approach reveals significant non-local effects on the transport of energetic $\alpha$-particles. This has a direct impact on hydrodynamic spatial profiles during combustion: the hot spot reactivity is reduced, while the inner dense fuel layers are preheated by the escaping $\alpha$-suprathermal particles, which are transported farther out of the hot spot. We show how the kinetic transport enhancement of fusion products leads to a significant reduction of the fusion yield. 
  
\end{abstract}

\maketitle
\section{Motivation and context of the study}
\label{sec:intro}
The design of Inertial Confinement Fusion (ICF) targets and the interpretation of ICF experiments rely on numerical simulations based on hydrodynamic Lagrangian codes where  kinetic effects are only considered as corrections included in the transport coefficients \cite{LIN981, Atzeni}. In particular, ion thermal conduction is treated approximately, at best through Spitzer-Braginskii local formulae \cite{SPI532,BRA65A}, and non-Maxwellian features in the ion velocity distributions are always neglected. Ion viscosity effects are also usually
not taken into account, in favor of numerical (``pseudo'') viscosity. This may have an impact on the modeling of shock waves that propagate throughout the capsule during implosion \cite{CAS91A, VID95A, LAR03A}, especially as the shock wave is reflected from the hot spot center.

More precisely, the fluid description is relevant if the mean free path of plasma particles, namely electrons and ions, is smaller than the characteristic length scale. Although this condition is reasonably fulfilled during the implosion stage for thermal deuterium and tritium (D,T) ions, it does not apply to fast particles, in particular to fusion products -suprathermal $\alpha$-particles- near the ignition threshold and during the combustion phase. Indeed, it is known \cite{FRA744} that the density of the central hot spot is such that  the mean free path $\lambda_\alpha$ of fast $\alpha$-particles is roughly equal to the hot spot radius $R$. 
 
Nevertheless, in all present-day fluid codes, multi-group flux limited diffusion schemes \cite{COR754,POM33} are usually employed to model suprathermal $\alpha$-particles. This kind of methods, although computationally efficient, relies on the assumption that the fast particle mean free path is smaller than the characteristic scale length of the energy deposition zone. Since this hypothesis does not hold for energetic particles in a typical ICF target, diffusion methods may not calculate the energy and momentum deposition associated to fast fusion products accurately. This may have significant consequences on the modeling of the ignition and combustion phases. 

Moreover, the \textit{coupling} between suprathermal particles and the thermal bulk is usually treated in a rough manner in fluid models, by removing the suprathermal particles that are slowed down below a given energy threshold and injecting the removed particles in the thermal bulk. Therefore, the thermalization process is not described with  sufficient precision. Those approximations may also influence the calculation of ignition thresholds and energy gains.

In this work, we present a full ion-kinetic modeling of suprathermal fusion products, treated self-consistently  with the ion-kinetic modeling of the thermal imploding plasma. The difficulty lies in the coupling of ion populations characterized by two different energy scales:
thermal $D$,$T$ ions, which form the bulk of the imploding plasma and whose kinetic energy is in the keV range, are coupled to suprathermal $\alpha$-particles, created at 3.52 MeV by fusion reactions. 

To overcome the difficulty associated to the high energy contrast, we develop a \textit{two-energy-scale approach} in which we consider the $\alpha$ distribution function as a set of two components ((namely a suprathermal and a thermal one) evolving on two well distinguished characteristic velocity scales. We  show that it is possible to rearrange the terms of the Fokker-Planck equation governing the evolution of fast fusion products, in such a way that that the collision operator is recast into a \textit{system} coupling two \textit{components}  associated to the $\alpha$ distribution function. Each component is associated to a particular velocity scale.

This strategy enables us to design tractable numerical methods, which have been employed to build a new ion kinetic code \textsc{Fuse} (for \emph{\textsc{Fpion} Upgrade with two Scales of Energy}) \cite{jcp2}, built as an extension of the former code \textsc{FPion} \cite{CAS91A}.  Note that existing ion kinetic codes can only describe the implosion of DT targets in sub-ignition conditions  \cite{VID95A, LAR03A}. Besides, the energy released from fusion reactions is not accounted for in a self-consistent manner.

A new computationally efficient approach developed recently \cite{jcp2} enables us to thoroughly study  ion kinetic effects during implosion, ignition and combustion stages of real ICF target configurations. 

The paper is organized as follows: in Sec.\ref{sec:theory}, we describe the theoretical model applied to the DT fuel. We present a hybrid electron fluid-ion kinetic model. First, we develop the Vlasov-Fokker-Planck (VFP) formalism for thermal species. Then, we extend the ion kinetic modeling to suprathermal $\alpha$-particles produced by fusion reactions. A physical analysis of the slowing down process shows how the \textit{two-component} nature of the $\alpha$ distribution function builds up. We then present an original two-velocity-scale decomposition of the Fokker Planck operator related to $\alpha$-particles. In particular, we explain how to accurately treat the coupling between the suprathermal particles and the thermal bulk. Then, in Sec.\ref{sec:application}, we compare a full ion kinetic simulation of a realistic ICF configuration with the fluid calculation, during the implosion, ignition and combustion processes. More precisely, we  consider a baseline 1D (\textsc{Fci1}) spherical fluid calculation of a typical ICF target. We then start a full ion kinetic calculation carried out with our code \textsc{Fuse} on the same design considered 1 ns before stagnation. This approach allows us to lay the emphasis on ion kinetic effects, especially during the end of the implosion process, as well as during the birth of the burn front and its subsequent propagation through the dense fuel shell.

\section{Description of the theoretical model}\label{sec:theory}

To develop a numerically tractable strategy to describe thermal species and energetic particles at a kinetic level, we go back to the underlying physical model described by the Vlasov Fokker Planck equations applied to the DT fusion plasma considered in typical ICF conditions. 



\subsection{Ion-Kinetic modeling of the thermal bulk}

The DT fuel is modeled by an hybrid electron-fluid/ion-kinetic approach \cite{DEG1}. The model is relevant since the characteristic time of the considered problem is close to the ion-ion collision time $\tau_{ii}$, which is significantly greater than the electron equilibrium time $\tau_{ee}$. More precisely, we have the following ordering\cite{BRA65A}: $\tau_{ee} \sim \varepsilon \tau_{ii}$, where $\varepsilon = (m_e/m_i)^{1/2} \sim 0.022$. As a consequence, at the relevant time scale $\tau_{ii}$, the electron kinetic equation reduces to a fluid equation.
Moreover, since $\tau_{ii}>>1/\omega_{pe}$, $\omega_{pe}$ being the electron plasma frequency, and the characteristic length is of the order of the ion collisional mean free path $\lambda_i>>\lambda_{De}$, $\lambda_{De}$ being the electron Debye length, the quasi-neutrality assumption is relevant. We then have:

\begin{equation}
\label{eq:quasineut}
n_e =\sum_i Z_i n_i, \qquad\vec V_e = \sum_i Z_i n_i \vec V_i,
\end{equation}
where the summation is carried out over all ion species (D,T and $\alpha$). $n_e$ (resp. $n_i$) denotes the electron (resp. ion) density and $\vec V_e$ (resp. $V_i$) refers to the mean flow electron (resp. ion) velocity.

Consequently, only an equation for the electron temperature (or, equivalently, the energy density) is actually needed since  the electron density and velocity are known from the quasi-neutrality conditions (\ref{eq:quasineut}).

The electron fluid model is coupled to a full ion-kinetic model, based on a Vlasov-Fokker-Planck formalism. More precisely,
each ion species $i$ (namely D,T and $\alpha$-ions) with atomic mass $A_i$ and charge $Z_i$ is described by a distribution function $f_i(r,\vec v,t)$. Considering a spherical one-dimensional geometry, $r$ designates the spatial radius and $\vec v$ the velocity.  Note that an azimuthal symmetry in velocity space around $v_r$ holds due to the assumed spherical symmetry in configuration space. The velocity vector can then be represented by a set of two coordinates $(v_r,v_\bot)$ such that $v=v_r \vec e_r + v_\bot \vec e_\bot$, or, equivalently, $(v,\theta)$ such that $\vec v= v\cos\theta \vec e_r + v\sin\theta \vec e_\bot$. 

Besides, to manipulate numbers that are closed to unity, physical quantities are expressed in the units defined on Table \ref{tabunit}. In particular, the normalized ion distribution function $f_i(r, \vec v)$ is defined from the dimensional distribution function $F_i(R,   \vec V)$ combining the corresponding reference thermal velocity $v_{i}^{th} \sim \sqrt{T_0/m_i}$ and the reference density $n_0$ in the following way:

\begin{equation}
f_i\left(r = \displaystyle\frac{ R}{\lambda_0},\vec v = \frac{V}{v_i^{th}}\right) = \displaystyle\frac{\left(v_i^{th}\right)^3}{n_0} F_i(R, \vec V)
\end{equation} 

\begin{table}[!h]
\caption{Units defined from reference values of the particle density $n_0$ and
particle thermal energy $T_0$.}
\label{tabunit}
\begin{center}
\begin{tabular}{l l}
\hline\noalign{\smallskip}
Quantity & Unit\\
\noalign{\smallskip}\hline\noalign{\smallskip}
density & $n_0$ (arbitrary reference value) \\
thermal energy & $T_0$ (arbitrary reference value) \\
time & $\tau_0 = T_0^{3/2}m_p^{1/2}/4\pi e^4n_0$ \\
length & $\lambda_0 = (T_0/m_p)^{1/2}\tau_0 = T_0^2/4\pi e^4n_0$ \\
velocity & $v_0 = (T_0/m_p)^{1/2} = \lambda_0/\tau_0$ \\
distribution function & $f_0 = n_0/v_0^3$ \\
first Rosenbluth pot. & $\mathcal{S}_0 = n_0/v_0$ \\
second Rosenbluth pot. & $\mathcal{T}_0 = n_0v_0$ \\
electric field ($\mathcal{E}_i$) & $\mathcal{E}_0=m_pv_0^2/\lambda_0=m_p\lambda_0/\tau_0^2$ \\
heat flux & $Q_0 = n_0T_0^{3/2}/m_p^{1/2}$ \\
\noalign{\smallskip}\hline
\end{tabular}
\end{center}
\end{table}

Distribution functions associated to thermal species (namely D,T ions and thermalized $\alpha$-particles) are expressed using a cylindrical velocity parametrization $(v_r,v_\bot)$. This choice guarantees a homogeneous accuracy in the whole velocity domain and enables us to model possible non-Maxwellian distributions with sufficient precision.

The thermal and normalized distribution function $f_i$ satisfies the following Vlasov-Fokker-Planck (VFP) equation which is expressed in dimensionless units as~:
\begin{eqnarray}\label{eq:vfp_ith}
&&\frac{\partial f_i}{\partial t}+ v_r\frac{\partial f_i}{\partial r} +
\frac{v_\bot}{r}\left(v_\bot\frac{\partial f_i}{\partial v_r}
- v_r\frac{\partial f_i}{\partial v_\bot}\right)+ \frac{\mathcal{E}_i}{A_i}\frac{\partial f_i}{\partial v_r}
 \nonumber\\
&& = \sum_{j=1}^n\left(\frac{\partial f_i}{\partial t}\right)_{ij} + \left(\frac{\partial f_i}{\partial t}\right)_{ie}
\end{eqnarray}

The last term on the left-hand side of Eq.(~\ref{eq:vfp_ith}) involves the effective electric field $\mathcal{E}_i$ which accelerates the ion species $i$. It is defined by~:

\begin{equation}\label{Ei}
\mathcal{E}_i = - (Z_i/n_e)\,\partial P_e/\partial r.
\end{equation}
where $P_e$ is the dimensionless electron pressure.

The expression of the electrostatic field (\ref{Ei}) can be obtained by developing the electron momentum conservation equation (corresponding to  the second moment of the electron Fokker-Planck equation) with respect to the ratio $\varepsilon$ and disregarding terms of order $\mathcal{O}(\varepsilon)$.

We then expand the collision terms that appear in the right-hand side of Eq.~(\ref{eq:vfp_ith}). In a fully ionized plasma such as the one considered here, large angle scattering is much less likely than the net large-angle deflection due to a cumulative effect of many small-angle collisions that the projectile experiences along its path \cite{ROS573}. Each of the collision terms in the right hand side of Eq.\,\eqref{eq:vfp_ith} can then be expressed  as a Fokker-Planck operator in velocity space, which amounts essentially to an advection-diffusion form. More precisely, the first term in the left-hand side of Eq.~(\ref{eq:vfp_ith}) models the collisions between ions and is given by the following Fokker-Planck form~:

\begin{equation}\label{eq:vfp_ith}
\left(\frac{\partial f_i}{\partial t}\right)_{ij} = \frac{4\pi Z_i^2 Z_j^2}{A_i^2}
\mbox{Log}\Lambda_{ij}\frac{\partial}{\partial v_\alpha}\left[\frac{A_i}{A_j}\frac{\partial
\mathcal{S}_j}{\partial v_\alpha} f_i -\frac{\partial^2 \mathcal{T}_j}{\partial
v_\alpha\partial v_\beta}\frac{\partial f_i}{\partial v_\beta}\right]
\end{equation}

where $\mathcal{S}_j$ and $\mathcal{T}_j$ are the so-called Rosenbluth potentials \cite{ROS573} associated to the target ions $j$. They are defined by a set of Poisson equations in velocity space:
\begin{equation}
\label{eq:poisson_rosenbluth}
\Delta_v \mathcal{S}_i = f_i, \qquad \Delta_v \mathcal{T}_i = \mathcal{S}_i.
\end{equation}

The Coulomb logarithm $\mbox{Log}\Lambda_{ij}$ (for any species $i,j$ including electrons) is related to the Coulomb potential screening  and taking quantum effects into account: $\Lambda_{ij}=\lambda_D/\max\{\lambda_{\rm bar},\rho_\bot\}$. The Debye length
$$ \lambda_D =\left(4\pi n_e e^2/T_e+\sum_{j=1}^n 4\pi n_j Z_j^2 e^2/T_j\right)^{-1/2} $$
depends on the temperature $T_j$, which is expressed in energy units. It is related to the thermal ion distribution function $f_j$  by the relation:
$$ T_j = \frac{m_j}{3n_j}  \int (v-V_j)^2 f_j(\vec v)\, d^3v,$$
where $n_j = \int f_j(\vec v)\, d^3 v$ is the density of ion species $j$ and $\vec V_j = n_j^{-1} \int \vec v f_j(\vec v)\, d^3v$ is their mean velocity. The characteristic lengths $\rho_\bot$ and $\lambda_{\rm bar}$ are the  classical and quantum impact parameters:
$$ \rho_\bot =  Z_a Z_be^2/m_{ij}u_{ij}^2, \qquad \lambda_{\rm bar} =  \hbar/m_{ij}u_{ij}$$
where $m_{ij}=m_i m_j/(m_i+m_j)$ is the reduced mass and $u_{ij}= \sqrt{3}(T_i/m_i + T_j/m_j)^{1/2}$ is an average relative velocity between the particle species $i$ and $j$. The Coulomb logarithm is thus a particular function of hydrodynamic quantities. It is symmetric with respect of particle species, $\Lambda_{ij}=\Lambda_{ji}$, which is related to the energy and momentum conservation during the collision.

The second term in the right hand side of (\ref{eq:vfp_ith}) models the effect of collisions between  thermal ion species $i$ and electrons. It is  expressed as another Fokker-Planck term, in which the electron distribution function is approximated by a Maxwellian function characterized by a density $n_e$, a mean velocity $\vec u_e$ and a temperature $T_e$:


\begin{equation}
\label{eq:fp_i_e}
\left.\frac{\partial f_i}{\partial t}\right|_{i e} = \displaystyle\frac{1}{\tau_{e i}} \frac{\partial}{\partial \vec v}\cdot \left[(\vec v - \vec u_e) f_i(\vec v) + \frac{T_e}{A_i} \frac{\partial f_i}{\partial v_i}(\vec v) \right],
\end{equation}
where $\tau_{ei }$ is a characteristic dimensionless $e-i$ collision time defined by:
\begin{equation}
\label{eq:tauei}
\tau_{e i} = \frac{3\sqrt{\pi}A_i T_e^{3/2}}{2\epsilon \sqrt{2}Z_i^2n_e\mbox{Log}\Lambda_{i e}}.
\end{equation}
Equation \eqref{eq:fp_i_e} is obtained by a truncated expansion of the full ion-electron Fokker-Planck operator with respect to the small parameter $\varepsilon$ \cite{CAS91A, LAR03A}.

\subsection{Two scale-kinetic modeling of suprathermal $\alpha$ particles}

The kinetic model applied to thermal particles can be extended to treat suprathermal $\alpha$-particles created by fusion reactions in a self-consistent manner.

Qualitatively, once created by fusion reactions, suprathermal $\alpha$-particles are transported through an inhomogeneous plasma and slowed down through  Coulomb collisions with the electrons and thermal ions. Besides, pressure gradients give rise to an electrostatic field  $\vec{\mathcal{E}}(\vec r,t)$ that may accelerate or decelerate $\alpha$-particles. To give an accurate description of the transport, as well as the non-local energy and momentum exchange that occur between $\alpha$-particles and the thermal bulk, a full kinetic modeling based on the Vlasov-Fokker-Planck equation is required.

The distribution function $f_\alpha(\vec r,\vec v,t)$ of $\alpha$-particles characterized by a dimensionless charge $Z_\alpha$ and a mass $A_\alpha$ is governed by the VFP equation~:

\begin{equation}
\label{eq:vfp_alpha} 
\displaystyle\frac{\partial f_\alpha}{\partial t}+\vec v \cdot \frac{\partial f_\alpha}{\partial \vec r} +
\frac{Z_\alpha \vec{\mathcal{E}_\alpha}}{A_\alpha}\cdot\frac{\partial f_\alpha}{\partial \vec v} =\sum_{i}\left.\frac{\partial
f_\alpha}{\partial t}\right|_{\alpha i} + \left.\frac{\partial f_\alpha}{\partial t}\right|_{\alpha e} + \left.\frac{\partial f_\alpha}{\partial t}\right|_{\rm fuse}.
\end{equation}

The fist two terms in the right-hand side of (\ref{eq:vfp_alpha}) models the effects of collisions between $\alpha$-particles with thermal ions and electrons. There are respectively  given by (\ref{eq:vfp_ith}) and (\ref{eq:fp_i_e}), where we set $i=\alpha$.  

The last term in \eqref{eq:vfp_alpha} represents the creation of $\alpha$-particles by fusion reactions. The source term is supposed to be isotropic and is given by:
\begin{equation}
\label{eq:evol_fdh_source}
\left.\frac{\partial f_\alpha}{\partial t}\right|_{\rm fuse}= \mathcal{R}_{DT}(\vec r,t)\frac{\delta(v-v_h)}{4\pi v^2},
\end{equation}
where $v_h= 1.3\times 10^9$\,cm.s$^{-1}$ is the initial velocity of suprathermal $\alpha$-particles whose initial energy is 3.52\,MeV. $\mathcal{R}_{DT}$ is the fusion reaction rate expressed as a function of the distribution functions of D and T, respectively:
\begin{eqnarray}
\label{eq:tau_reac}
&& \mathcal{R}_{DT}(\vec r,t) = \int\int f_D(\vec r,\vec v_D,t)\, f_T(\vec r,\vec v_T,t)\,|\vec v_D - \vec v_T|\,\nonumber\\
&& \times\sigma_{DT}(|\vec v_D - \vec v_T|)\,d^3v_D d^3v_T.
\end{eqnarray}
The distribution functions $f_D$ and $f_T$ are solutions of the Vlasov-Fokker-Planck equations  written for the deuterium and tritium species, respectively. Since $f_D$ or $f_T$  are not necessarily Maxwellian functions, possible Non-Local-Thermodynamic-Equilibrium (Non-LTE) effects in the tails of the distribution functions (which contribute mainly to the hot spot reactivity) are naturally taken into account. Integrals in Eq.\,\eqref{eq:tau_reac} are taken over the three-dimensional velocity space.

Let us briefly recall\cite{jcp2} how the two-component feature of the $\alpha$ distribution function builds up. 


It is known \cite{FRA744} that the beginning of the slowing-down of suprathermal $\alpha$-particles is governed nearly exclusively by electrons. The first stage of the $\alpha$ slowing down is thus described by:
\begin{equation}
\label{eq:fsure}
\left.\frac{\partial f_\alpha}{\partial t}\right|_{\rm coll} = \frac{1}{\tau_{\alpha e}} \frac{\partial}{\partial \vec v}\cdot \left[(\vec v - \vec u_e) f_\alpha(\vec v) + \frac{T_e}{m_\alpha} \frac{\partial f_\alpha}{\partial \vec v}(\vec v) \right].
\end{equation}
As long as $v \gg u_e$, the dynamic friction term (first term in the right hand side of (\ref{eq:fsure})) dominates so that the $\alpha$ distribution evolves with respect to:
\begin{equation}
\label{eq:fsurebis}
\left(\frac{\partial f_\alpha}{\partial t}\right)_{coll}  \approx \frac{1}{\tau_{\alpha e}}  \frac{1}{v^2}\frac{\partial}{\partial v}\cdot \left[v^3  f_\alpha(v) \right].
\end{equation}
The stationary solution of Eq.~(\ref{eq:fsurebis}) behaves as $f_\alpha \sim 1/v^3$, where $v$ is the suprathermal $\alpha$-particle velocity. Consequently, as long as fast $\alpha$-particles remain far from the thermal velocity domain, their distribution function varies smoothly over the whole suprathermal velocity region. 


Then, when slowed down $\alpha$-particles get closer to the thermal domain but still remain suprathermal, thermal ions  tend to dominate the end of the relaxation process, which is then governed by the equation:
\begin{equation}
\label{eq:modele_1d}
\left.\frac{\partial f_\alpha}{\partial t}\right|_{\rm coll} =\sum_{i} 4\pi\Gamma_{\alpha i}\frac{\partial}{\partial\vec v}\cdot\left(\frac{m_\alpha}{m_i}f_\alpha \frac{\partial\mathcal{S}_i}{\partial \vec v}\right),
\end{equation}
where only the dynamical friction term is retained for the present discussion. We have introduced:

$$
\Gamma_{\alpha i}= (4\pi Z_\alpha^2 Z_\beta^2/A_i^2)\mbox{Log}\Lambda_{\alpha i}.
$$

\begin{figure}[!h]
\centering
{\includegraphics[width=0.4\textwidth]{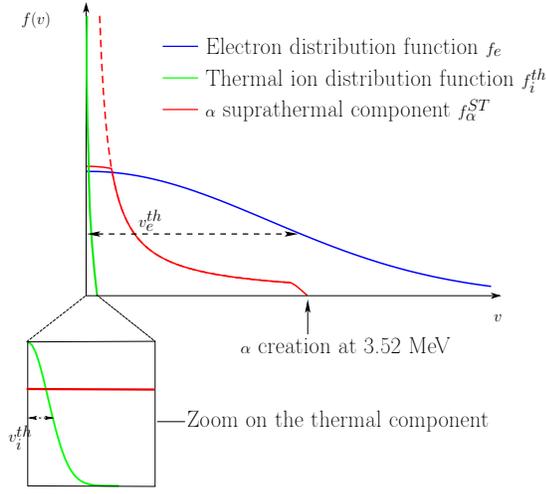}}
\caption{\label{fig:twoscale} Schematic representation of the collisional relaxation of suprathermal $\alpha$-particles on thermal target ions $i$. The red broken line refers to the stationary solution of Eq. (\ref{eq:fsurebis}).
}
\end{figure}

Qualitatively, one can consider that the distribution function of the thermal target species $i$, appears for suprathermal $\alpha$-particles highly localized in velocity space (see Fig.~\ref{fig:twoscale}). 


Besides, the divergence with respect to velocity that appears in the right hand side of Eq.\,\eqref{eq:modele_1d} can be expanded as follows:
$$
\frac{\partial}{\partial \vec v}\cdot \left(\frac{\partial\mathcal{S}_i}{\partial \vec v}f_\alpha\right)\simeq \frac{\partial\mathcal{S}_i}{\partial \vec v}\cdot \frac{\partial f_\alpha}{\partial \vec v} + f_\alpha \Delta_v \mathcal{S}_i.
$$
Using the approximation $ f_i(\vec v) = n_i \displaystyle\frac{\delta(v)}{4\pi v^2}$, which is valid for suprathermal $\alpha$-particles, the first Rosenbluth potential associated to the target ions $i$ can be calculated explicitly: $\mathcal{S}_i(v) \sim -n_i/4\pi v$. Then, by calculating its derivative, the slowing down of $\alpha$ particles can be modeled by:
\begin{equation}
\label{eq:modele_1dbis}
\left.\frac{\partial f_\alpha}{\partial t}\right|_{\rm coll} =\sum_{i} 4\pi\Gamma_{\alpha i} \frac{m_\alpha}{m_i}\left( \frac{\partial f_\alpha}{\partial \vec v} \cdot \frac{n_i}{4\pi v^2}\vec{e}_v + f_\alpha f_i  \right).
\end{equation}
The two terms in the right hand side of Eq.(\ref{eq:modele_1dbis}) have a clear physical sense. The first term $\sim \partial f_\alpha/\partial \vec v$ varies slowly and smoothly far from the thermal velocity domain. It can be characterized by a suprathermal velocity scale $v_{\alpha}^{ST}$, which is greater than the typical thermal ion velocity $v_i^{th}$. Actually, the term $\sim \displaystyle\frac{n_i}{4\pi v^2} \displaystyle\frac{\partial f_\alpha}{\partial \vec v}$ represents a conservative convection towards $v=0$. The associated convective rate $\displaystyle\frac{n_i}{4\pi v^2}$ increases as $v$ tends to $0$ so that the solution of:
\begin{equation}
\label{eq:convec_mod}
\left(\frac{\partial f_\alpha}{\partial t}\right)_{coll} =\sum_{i} 4\pi\Gamma_{\alpha i} \frac{m_\alpha}{m_i}\left[ \frac{\partial f_\alpha}{\partial \vec v} \cdot \frac{n_i}{4\pi v^2}\vec{e}_v \right]
\end{equation}
tends to a constant $f_0$ corresponding to the stationary state of (\ref{eq:convec_mod}). The part of the $\alpha$ distribution driven by (\ref{eq:convec_mod}) tends to be stretched and smoothed out as it approaches the thermal velocity region.

The second term $\sim f_\alpha f_i$ appears highly localized in the thermal region of velocity space and behaves qualitatively as a $\delta$-function for suprathermal $\alpha$-particles. This term  actually leads to the formation of a condensate of width $v_i^{th} \ll v_{\alpha}^{ST}$.


It thus seems natural to write the $\alpha$ distribution function as follows:
\begin{equation}
\label{eq:split_fd}
f_\alpha(\vec v,t) =  f_\alpha^{ST}(\vec v,t)+f_\alpha^{T}(\vec v,t),
\end{equation}
where: $f_\alpha^{ST}$ denotes the suprathermal component. It is defined on a large velocity domain, spreading to the MeV range; $f_\alpha^{T}$ is the thermal component. It is localized in the region of velocity space corresponding to target thermal ion distribution functions and vanishes in the suprathermal velocity domain. The original Fokker-Planck operator given in Eq.(\,\eqref{eq:modele_1d}) is then transformed into a \textit{system of two coupled equations} governing the two components $f_\alpha^{ST}$ and $f_\alpha^{T}$, respectively:
\begin{eqnarray}
\label{eq:system_ST_T}
 \left.\partial_t f_\alpha^{ST}\right|_{\alpha i} &=&  \Gamma_{\alpha i} \frac{n_i}{v^2} \partial_{v} f_\alpha^{ST} - n_i\Gamma_{\alpha i} f_\alpha^{ST} \frac{\delta(v)}{v^2}, \nonumber\\
 \left.\partial_t f_\alpha^T\right|_{\alpha i} &=& 4\pi\Gamma_{\alpha i} \partial_{\vec v}\cdot\left(f_\alpha^{T} \partial_{\vec v}\mathcal{S}_i\right) \nonumber\\
& & + 4\pi\Gamma_{\alpha i}  f_i  f_\alpha^{ST}(v=0).
\end{eqnarray}
This re-arrangement of the terms of the Fokker Planck operator enables us to design tractable numerical methods \cite{jcp2} treating the coupling between the two energy scales efficiently. Note that the coupling function between the two components takes two particular forms, depending on the considered velocity scale:
\begin{itemize}
\item{For the suprathermal component, we have $f_\alpha^{ST} f_i \sim n_i f_\alpha^{ST} \displaystyle\frac{\delta(v)}{4\pi v^2}$  since thermal target ions appear highly localized.}
\item{For the the thermal component, we can consider $f_\alpha^{ST} f_i \sim f_\alpha^{ST}(0) f_i$ since the suprathermal component is almost constant on the thermal velocity scale $v_i^{th}$. The term $\sim f_\alpha^{ST}(0) f_i$ appears as a source term for the thermal component. It corresponds to a feeding by the suprathermal component.}
\end{itemize}


Note that this two-component description of the $\alpha$ distribution function remains relevant if the velocity threshold $v_c$ where the thermal ions tend to dominate the $\alpha$ slowing down is much greater than the thermal target velocity scale $v_i^{th}$. In Fig.~\ref{fig:twoscale}, 
the threshold velocity $v_c$  corresponds  to the velocity at which the suprathermal component diverges from the solution of Eq. (\ref{eq:fsurebis}) modeling the pure effect of $\alpha$-electron collisions. For $v<v_c$, the suprathermal component is cut off by the effect of $\alpha-i$ collisions and tends to become constant as it approaches the thermal velocity region. An estimate of $v_c$ is given by the relative importance of electrons and ions on the slowing down of $\alpha$-particles. By retaining only the dynamical friction terms from the corresponding Fokker-Planck operators, the ratio $R_{i/e}$ between the ion and electron collisional drag can thus be approximated by~:
$$ R_{i/e} = \left.\frac{\partial f_\alpha}{\partial t}\right|_{\alpha i}\left/\right.\left.\frac{\partial f_\alpha}{\partial t}\right|_{\alpha e} \sim \frac{T_e^{3/2}}{v^3 m_e^{1/2}m_i} \sim \frac{T_e^{3/2}}{v^3 m_i^{3/2} \epsilon}. $$

The ration $R_{i/e}$ thus involves a characteristic threshold velocity $v_c$ defined by~: 
\begin{equation}
\label{eq:vcoup}
v_{c}= \epsilon^{-1/3}(T_e/m_i)^{1/2} \sim  3.6 \,v_i^{th}.
\end{equation} 

The condition $v_c >> v_i^{th}$ is thus reasonably fulfilled, so that the two-component description is relevant to model $\alpha$-particles in plasma conditions corresponding to the ignition and burn stages.

\subsection{Full ion kinetic model for thermal and suprathermal species}

The full VFP model of $\alpha$-particles is thus given by a set of two coupled equations. The equation for the fast $\alpha$-particles reads~:

\begin{eqnarray}
&& \frac{\partial f^{ST}_\alpha}{\partial t} +  v\,\cos\theta\,\frac{\partial f^{ST}_\alpha}{\partial r} + \frac{ 
\mathcal{E}_\alpha}{A_\alpha} \cos\theta \frac{\partial  f^{ST}_\alpha}{\partial v} \nonumber\\
&& = \sum_{i}  \Gamma_{\alpha i} \displaystyle\frac{\partial}{\partial \vec v} \cdot \, \left[\frac{n_i}{ v^2} \left(\frac{A_\alpha}{A_i} f_\alpha^{ST} \vec{e}_v + \frac{1}{2}\frac{\partial f_\alpha^{ST}}{\partial\theta}\vec{e}_\theta \right)\right] \nonumber \\
&&  + \frac{1}{\tau_{e\alpha}}\displaystyle\frac{\partial}{\partial \vec v} \cdot \,\left[(\vec v - \vec {u_e})f_\alpha^{ST} + \frac{T_e}{A_\alpha}\frac{\partial}{\partial \vec v}f_\alpha^{ST}\right] \nonumber \\
&&- \sum_{i=D,T,\alpha} 4\pi\Gamma_{\alpha i}\frac{A_\alpha}{A_i} f^{ST}_\alpha f_i^T \nonumber\\
&& + \mathcal{R}_{DT}(\vec r,t)\frac{\delta(v-v_h)}{4\pi v^2},
\label{eq:FP_supra_discr}
\end{eqnarray}

We choose a polar parametrization of the suprathermal $\alpha$ distribution function as $f^{ST}_\alpha(r,v,\theta,t)$, where two velocity components $(v,\theta)$ are such that $\vec v = v \cos\theta \,\vec{e}_r + v\sin\theta \, \vec{e}_\bot$.
Indeed, the collision term between suprathermal $\alpha$-particles and ions takes  a simpler form expressed in polar coordinates. In particular, the slowing down currents are co-linear with the local polar basis vectors $\vec{e}_v,\vec{e}_\theta$ of the velocity space. Such a parametrization choice facilitates the numerical resolution of Eq.(~\ref{eq:FP_supra_discr}).
We recall that the third term $\sim f_\alpha^{ST} f_i^T$ in the right-hand side of (\ref{eq:FP_supra_discr}) model the coupling with the thermal component from the suprathermal point of view. 

As energetic $\alpha$-particles slow down and reach the thermal velocity domain, a thermalized component $f_\alpha^T(r,v_r,v_\bot)$ builds up. The VFP equation governing the time evolution of the $\alpha$ thermal component is then~: 
\begin{eqnarray}
&& \frac{\partial f_\alpha^T}{\partial t} + v_r\frac{\partial f_\alpha^T}{\partial r} +
\frac{v_\bot}{r}\left(v_\bot\frac{\partial f_\alpha^T}{\partial v_r}
- v_r\frac{\partial f_\alpha^T}{\partial v_\bot}\right) +
\frac{\mathcal{E}_\alpha}{A_\alpha}\frac{\partial f_\alpha^T}{\partial v_r} \nonumber\\
&& = \sum_{i} 4\pi\Gamma_{\alpha i} \displaystyle\frac{\partial}{\partial \vec v} \cdot \,\left(\frac{A_\alpha}{A_i} f_\alpha^T \frac{\partial\mathcal{S}_i}{\partial \vec v}  - \nabla^2\mathcal{T}_i\frac{\partial f_\alpha^{T}}{\partial\vec v} \right) \nonumber \\
&& \qquad\qquad + \frac{1}{\tau_{e\alpha}}\displaystyle\frac{\partial}{\partial \vec v} \cdot \,\left((\vec v - \vec {u_e})f_\alpha^T + \frac{T_e}{A_\alpha}\frac{\partial}{\partial \vec v}f_\alpha^T\right) \nonumber\\
&& \qquad\qquad +\sum_{i} 4\pi\Gamma_{\alpha i}\frac{A_\alpha}{A_i} f^{ST}_\alpha f_i^T .
\label{eq:eqFP_t}
\end{eqnarray}

The thermal component velocity dependence is parametrized with respect to cylindrical coordinates $v_r,v_\bot$, since it evolves on the same velocity mesh as the other thermal ion species (D,T).
The source term coming from the slowing down of the suprathermal component appears in the last term of the right-hand side of \eqref{eq:eqFP_t}. As far as the thermal component is concerned, the suprathermal component $f^{ST}_\alpha$ appears relatively constant over the whole thermal velocity grid since it varies significantly on the coarse suprathermal velocity grid whose mesh size is of the order of the thermal velocity. That is why we can use the following estimate~:
\begin{equation}
\label{eq:source_term_T}
\sum_{i} 4\pi\Gamma_{\alpha i}\frac{A_\alpha}{A_i} f^{ST}_\alpha f_i^T \sim f^{ST}_\alpha (V_0)\sum_{i} 4\pi\Gamma_{\alpha i}\frac{A_\alpha}{A_i}  f_i^T .
\end{equation}

This procedures guarantees an exact mass conservation: the number of particles that are removed from the suprathermal component are injected into the thermal component. Besides, summing Eqs. (\ref{eq:eqFP_t}) and (\ref{eq:FP_supra_discr}) gives the original Fokker-Planck operator Eq.\,\eqref{eq:modele_1d}, so that the splitting method presented here preserves each moment associated to the $\alpha$ distribution function.

\section{Kinetic simulation of the implosion and combustion of an ICF ignition target}\label{sec:application}

In this section, we compare the kinetic modeling with the fluid approach where a multi-group diffusion scheme is applied to simulate $\alpha$-particles.

We consider a baseline  1D spherical fluid simulation of the implosion of an ICF
target carried out with the hydrodynamic code \textsc{Fci1}. The chosen  parameters are typical of ignition capsules designed for the \textsc{Nif} \cite{LIN014}, namely $0.25$ mg of cryogenic DT deposited on the inner
surface of a CH shell of 1 mm (inner) radius. 
The main features of the considered fluid calculation are the following:
\begin{itemize}
\item{the maximum areal density of the fuel is 1.7 g.cm$^{-2}$, reached at the time $t_1^{f} = 18340$ ps, after the beginning of the implosion process.}
\item{The beginning of the combustion process starts around that time, and the fusion reaction power rises to $P_{th} \sim 2.5\times 10^{18}$ W, at the time $t_2^{f} = 18360$ ps.} 
\item{The total energy released by fusion reaction is 19 MJ.}
\end{itemize}

On the other hand, the kinetic calculation carried out with our code \textsc{Fuse} is started at $t=17.34$ ns, before the main converging shock
reaches the center of the target. This time corresponds approximately to 1 ns before
the stagnation and burn as predicted by the hydrodynamic code for the considered ICF ignition target.

The implosion is driven by a boundary condition which is taken from the
hydrodynamic quantities recorded as a function of time on the fuel/pusher interface in the fluid simulation . Note that when the burn front reaches the DT fuel boundary, the boundary condition coming from the fluid calculation may not be consistent with the kinetic calculation. The blowing off of the DT fuel modeled with our kinetic numerical approach  is then calculated after stagnation/ignition in a self-consistent manner with the total pressure evaluated at the external radius of the system. To simulate the subsequent blowing off of the system, we need an estimate of the pusher remaining mass, which is chosen to reproduce a dislocation that is consistent with the one calculated in the fluid calculation. We will control this approximation by studying the sensitivity of the kinetic calculation with respect to the average pusher mass.

The kinetic simulation considers three ion species, namely D, T and $\alpha$. Initially, only thermal species D and T are present. They give birth to suprathermal $\alpha$  particles by fusion reactions. The relaxation of the suprathermal $\alpha$  component then leads to the creation of an $\alpha$ thermal component interacting with the other thermal ion distribution functions (D and T, respectively). Note that the thermal bulk is described in more detail than in \cite{LAR03A} where a single mean ion species with a mass number of 2.5 was considered. 

In our kinetic simulation, the position of each spatial meshes is updated after each time step with respect to the imposed boundary condition and to the fixed number of spatial meshes $i_{\max}$. This updating is  performed before each advection phase. This means that the position of a given spatial cell $r_{i_0}$, with $1\leq i_0 \leq i_{\max}$ is time dependent, decreasing with the size of the imploding system.  To represent in a satisfactory manner both the dense region where the fluid simulation grid is the finest and the central zone where it is rather coarse, we employ 78 cells with a geometrically varying mesh size (with the ratio 0.97) so that the mesh size $\delta r$ is decreasing from 20\,$\mu$m near the center to less than one micron near the outer boundary. The thermal velocity space $(v_r,v_\bot)$ is discretized into $129\times 64$ cells, whereas the suprathermal velocity grid $(v,\theta)$ makes use of $100\times60$ cells. The reference time-step value is 0.05\,ps.
 
We make sure that kinetic simulations have reached convergence with respect to the spatial mesh, the velocity mesh (for each component) and the chosen time step.

Let us compare the fluid and kinetic simulation results, focusing on the combustion phase.


\subsection{Integrated DT fuel performances}

\begin{figure}[!h]
\centering
{\includegraphics[width=0.45\textwidth]{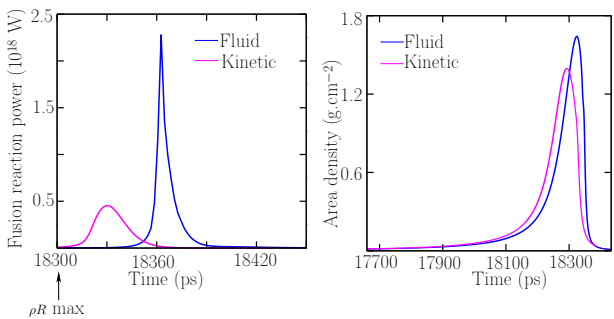}}
\caption{\label{fig:ptht_ror} Comparison of the power generated by fusion reactions (left) and total fuel  areal density (right) between the fluid and the kinetic model. 
}
\end{figure}

We firstly compare the integrated performances of the fuel, by plotting in Fig.~\ref{fig:ptht_ror} the time evolution of the power released by fusion reactions, as well as the time evolution of the total areal mass $\int \rho dr$ of the DT fuel. It can be seen that the global performances of the considered target are greatly reduced in the kinetic calculation. In particular~:

\begin{itemize}
\item{The maximum  power released by fusion reactions is more than 40 times less than in the fluid calculation.  The width of the power curve - related to the combustion time - is also larger: $\approx 20$ ps compared to $\approx$ 5 ps in the fluid modeling. It induces a significant reduction of the thermonuclear yield: only 11 MJ ($\pm 20 \%$) are released in the kinetic calculation, compared to 19 MJ in the reference fluid simulation.}
\item{The combustion starts also slightly earlier ($\approx 25$ ps) in the kinetic calculation, as one can infer from the fusion power time evolution. The time difference is relatively small, about 25-30 ps (compared to the simulation time $\sim$ 1000 ps).}
\item{The time behavior of the total areal mass of the fuel is consistent with the thermonuclear power evolution: $\int \rho_{DT} dr$ rises faster in the kinetic calculation, but reaches a smaller maximum value: $\rho R^{max} \sim 1.4$ g.cm$^{-2}$, compared to $\sim 1.7$ in the fluid model. The kinetic fuel burnt fraction is thus lower, which is consistent with the calculated yield reduction. } 
\end{itemize}

\begin{figure}[!h]
\centering
{\includegraphics[width=0.45\textwidth]{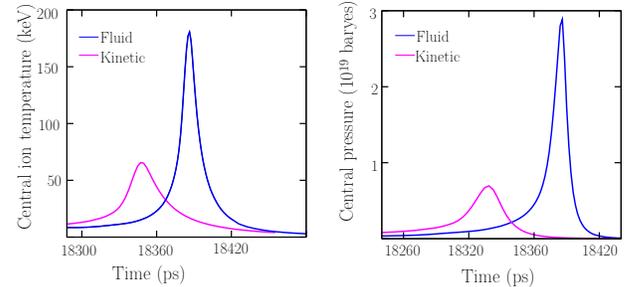}}
\caption{\label{fig:tipi_centre} Comparison of the central hot spot conditions (ion temperature (left) and total pressure (right)) between the fluid and the kinetic model.  
}
\end{figure}

The fuel integrated performance analysis is confirmed by the study of the central hot spot evolution. In particular, we plot the ion temperature and the central total pressure  as a function of time in Fig. \ref{fig:tipi_centre}. The hot spot reaches temperatures that are significantly lower in the kinetic calculation. The central pressure follows the same dynamic as the central hot spot temperature and is significantly lower in the kinetic calculation. Both quantities seem to reach the ignition threshold earlier in the kinetic modeling, which is consistent with the time evolution of $P_{th}(t)$ and $\int \rho dr$.

The significant differences observed on the global performances of the target between the kinetic and fluid model can be explained by identifying mainly two stages, that we will study more closely in the following:
\begin{itemize}
\item{During the implosion process, as the main shock reaches the center of the hot spot, non-collisional features start developing in the shock wave central region, so that the kinetic model looses synchronism with the fluid calculation. The shock front is broader and the central hot spot temperature rises earlier in the kinetic calculation. }
\item{This earlier heating of the hot spot leads  to an earlier ignition  that occurs in a bigger and less dense fuel in the kinetic modeling.  Besides, the full kinetic modeling of energetic $\alpha$-particles reveals a burn front structure that differs significantly from the fluid model (see for instance Fig.\ref{fig:burn_front}). The kinetic flame is in particular less localized than in the fluid model. We will show that this effect is related to the non-local transport of energetic $\alpha$-particles which deposit their energy and momentum as they accumulate in a region located out of the hot spot, at the entrance of the dense fuel shell.}
\end{itemize}


\subsection{Sensitivity to the mean pusher mass}

\begin{figure}[!h]
\centering
{\includegraphics[width=0.5\textwidth]{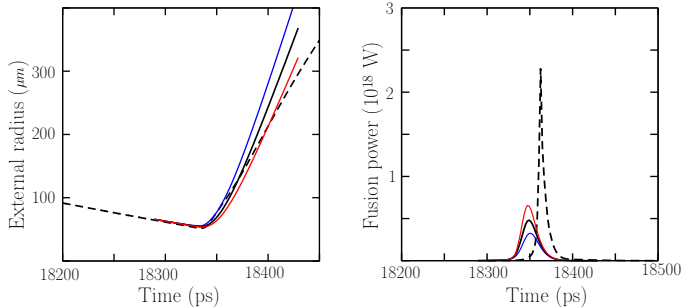}}
\caption{\label{fig:comp_ray_pmax} Fuel dislocation (left)  and total thermonuclear power (right) released by fusion reactions for different pusher masses in the kinetic calculation. The baseline hydrodynamic calculation is represented in black dashed line and the reference kinetic simulation correspond to the plain black curve. Red (resp. blue) curves correspond to a kinetic calculation where the average pusher mass is multiplied (resp. divided) by two. The implied variation on the fusion yield is approximately 20 $\%$. 
}
\end{figure}

When the burn front reaches the DT fuel boundary, the boundary condition coming from the fluid calculation may not be consistent with the kinetic calculation. The blowing off is driven by the total pressure evaluated at the external radius of the fuel. The pressure tends to set in motion the remaining part of the pusher, which is represented by an average inertia mass, that we choose to reproduce the same dislocation as in the fluid simulation. We are aware that this rather crude treatment may not be sufficient to accurately model the interaction between the dense fuel shell and the pusher. Nonetheless, we are interested in the development of the combustion inside the fuel, so that the modeling of the pusher by a mean inertia mass is a first approximation relevant for our purpose. To check the validity of this approximation, we carry out a full kinetic calculation of the combustion process with different equivalent pusher masses. In Fig.\ref{fig:comp_ray_pmax}, we plot the temporal evolution of the DT fuel external radius and the associated total thermonuclear fusion power. We can see  that the fusion yield is not too sensitive to the choice of the mean pusher mass. The fusion power curves represented in  Fig.\ref{fig:comp_ray_pmax}-right for different pusher masses remain in the same order of magnitude. The corresponding fusion energy variation is approximately 20$\%$. This approximation is then sufficient to describe the kinetic effects on the development of the combustion process inside the DT fuel. Studying in more detail the dislocation phase, in particular the interaction between the DT shell and the CH pusher, may require a tailored extension to our code, and will be addressed in a future work.

\subsection{Kinetic effects during implosion}

\begin{figure}[!h]
        \centering
   \includegraphics[width=0.5\textwidth, angle=0]{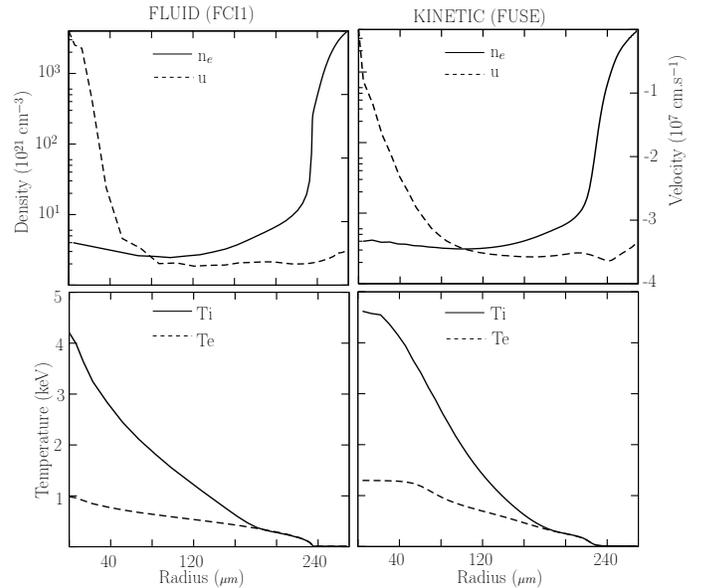}
   \caption{\label{hydr100}Comparison of the fluid(\textsc{Fci1}) (left) and kinetic(\textsc{Fuse}) (right) calculations. Profiles of the density, velocity (top panels) and of the electron and total ion temperatures (bottom panels)  in a DT ignition target  at the time $t=17.65$ ns, which corresponds to 310\,ps after the beginning of the kinetic calculation and roughly 650\,ps before the target ignition.}
\end{figure}

\begin{figure}[!h]
        \centering
   \includegraphics[width=0.5\textwidth, angle=0]{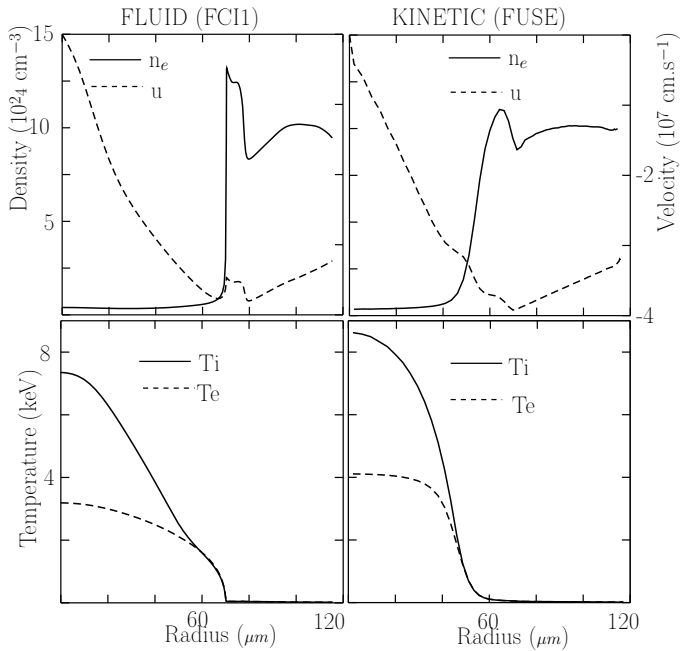}
   \caption{\label{hydr650}Comparison of the fluid(\textsc{Fci1}) (left) and the kinetic(\textsc{Fuse}) (right) calculations. Profiles of the density, velocity (top panels) and of the electron and total ion temperatures (bottom panels) in a DT ignition target  at the time $t=18.12$ ns, which corresponds to 780 \,ps after the beginning of the kinetic calculation. This is close to the time of the target stagnation and near the beginning  of ignition.}
\end{figure}

To study ion-kinetic effects during implosion,  it is instructive to compare the density, velocity and temperature spatial profiles calculated by the full kinetic simulation (carried out with our two-velocity-scale kinetic code \textsc{Fuse}) with those calculated by the hydrodynamic code \textsc{Fci1} at two different times of the implosion process:
\begin{itemize}
\item{at $t=17.65$ ns, that is to say 310 ps after the beginning of implosion, we find a good agreement between the \textsc{Fuse} kinetic calculation and the \textsc{Fci1} fluid simulation (see Fig.~\ref{hydr100}). Kinetic effects do not play a significant role at this stage.}
\item{At $t=18.12$ ns, in the vicinity of the target stagnation, the results of \textsc{Fuse} and \textsc{Fci1} are still in  relatively good agreement (see Fig.~\ref{hydr650}). However, we note that the compression zone near the inner interface of the dense fuel lies closer to the target center in the kinetic calculation (see the negative velocity gradient region about r = 70 $\mu$m in the right panel of Fig.~\ref{hydr650}). This feature has already been observed with \textsc{Fpion}\cite{LAR03A}. This is related to a perturbation of the ion heat flux by kinetic effects, especially at the interface between the hot spot and the dense shell. We will study in more detail this effect in the following.   Besides, the central hot spot temperature tends to rise more quickly in the kinetic modeling.  This apparent loss of synchronism between both simulations  tends to build up during the end of the implosion process.
}
\end{itemize}

\begin{figure}[!h]
\centering
{\includegraphics[width=0.5\textwidth]{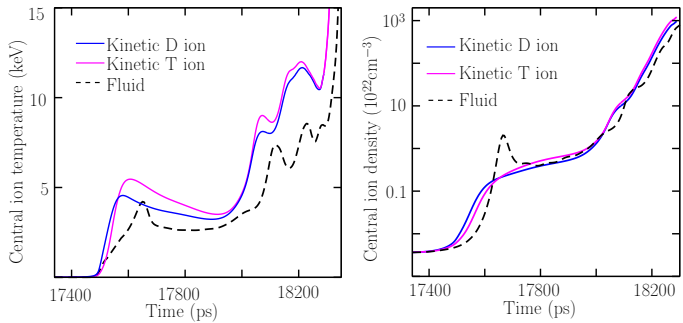}}
\caption{\label{fig:tite_implo} Comparison of the central hot spot ion temperature (left) and density (right) during implosion between the fluid and the kinetic model.}
\end{figure}

We now study in more detail the time evolution of the central ion temperature. A fluid/kinetic comparison  is shown in Fig.~\ref{fig:tite_implo}-left. It reveals significant discrepancies between the fluid approach and the full ion-kinetic treatment. In particular, one notices clear differences when the shock is reflected from the center of the hot spot. The temperature variations implied by the arrival of the shock wave in the central region appears slightly earlier in the kinetic implosion. Those temporal variations are also clearly broader and less sharp than in the fluid simulation. Thus, starting form the same initial state, corresponding approximately to 1 ns before ignition, both calculations  progressively loose synchronism in the such a way that the hot spot ion temperature tends to rise earlier in the full ion-kinetic modeling. 

This loss of synchronism is connected to a different modeling of ion shock structures: indeed, it is known \cite{CAS91A,MANH07} that ion-kinetic effects affect the structure of the shock wave propagating through the capsule, thus influencing the width of the front and the shock reflection from the hot spot center. More precisely, the kinetic shock width can be estimated by $\sim (m_i/m_e)^{1/2} \lambda_{ii}$, which could significantly be greater than the shock width calculated by  the fluid simulation where the ion front may be artificially more localized due to numerical pseudo-viscosity effects. As a consequence, the central temperature variations induced by the arrival of the shock waves  tends to take place earlier in the kinetic simulation and in a significantly wider region than the shock width of the fluid model.

To illustrate the significance of kinetic effects during implosion, we study the spatial profiles of the discrepancy $T_{\|}-T_{\bot}$ between the parallel and transverse ion temperatures. Those quantities are particular moments of the ion distribution function and are defined for a ion species $i$ by~:

$$
T^i_{\|} = \frac{m_i}{2}\int (v_r-u_i)^2 f_i(v_r,v_\bot) d^3 v,
$$
and

$$
T^i_{\bot} = \frac{m_i}{2}\int v_\bot^2 f_i(v_r,v_\bot) d^3 v.
$$

\begin{figure}[!h]
\centering
{\includegraphics[width=0.5\textwidth]{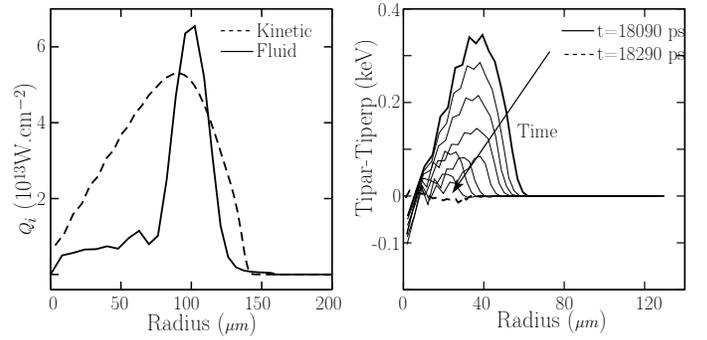}}
\caption{\label{fig:nlte_implo} Estimated deviation from the Maxwellian distribution function in the hot spot and enhanced ion heat flux in the central hot spot. On the left panel, one represents the ion heat flux spatial profile at $t=17800$ ps, corresponding to the arrival of the shock wave in the central region. On the right panel, one gives the spatial profiles of the discrepancy $T_{\|}-T_{\bot}$ (associated to ion viscosity effects) at different times during the end of implosion. }
\end{figure}

The total ion parallel/transverse temperature are then obtained by taking the average over the different ion species:

$$
T_{\|,\bot} = \frac{1}{n_e} \sum_i n_i T^i_{\|,\bot}.
$$
The corresponding spatial profiles during the end of the implosion process are represented in Fig.~\ref{fig:nlte_implo}-left.
Kinetic effects are clearly visible as the shock wave reaches the central region of the hot spot, where $T_{\|} > T_{\bot}$. This gives an illustration of ion viscosity effects, which are not precisely modeled in a fluid code, and may impact the propagation of the shock wave through the hot spot, especially during its reflection from the center. 
At latter times, collisional effects intensify and the discrepancy between  $T_{\|}$ and $T_{\bot}$ tends to disappear. Ion distribution functions are then closed to Maxwellian functions. 

Besides, those kinetic effects have a direct impact on the ion-heat flux in the DT fuel and thus influence the hot spot temperature evolution. From the known thermal ion-distribution functions, one computes the total ion-heat flux for a given thermal ion species $i$:
$$
Q_i= Q^i_{\|} +Q^i_\bot 
$$ 

where:
$$
Q^i_{\|} = \frac{m_i}{2}\int (v_r-u_i)^3 f_i(v_r,v_\bot) d^3 v,
$$
and

$$
Q^i_{\bot} = \frac{m_i}{2}\int (v_r-u_i)v_\bot^2 f_i(v_r,v_\bot) d^3 v.
$$

The total ion heat flux is then calculated by taking the average over the different ion species~:
$$
Q = \frac{1}{n_e} \sum_i n_i Q_{i}.
$$

The non-LTE effects observed in the center and close to the interface with the dense shell have a direct influence on ion heat flux $Q_i$ spatial profiles during implosion (see Fig.\ref{fig:nlte_implo}-left, corresponding to the time $t=17800$ ps). The kinetic ion heat flux is significantly higher in the central hot spot, thereby explaining why the central ion temperature tends to rise faster during implosion. The kinetic ion heat flux also deviates from the fluid one in the spatial region corresponding to the interface between the dense fuel and the hot spot. The ion heat flux is also less peaked and broader in the kinetic modeling, due to non-local ion transport effects. This explains why the hot spot/dense shell interface tends to lie closer to the center in the kinetic model (see Fig.~\ref{hydr650}).

Moreover, in the multi-species kinetic calculation considered here, the T species tends to be slightly hotter than the D species in the central hot-spot region during implosion (Fig.\ref{fig:tite_implo}-left). The discrepancy between D and T ion central temperatures $T_T -T_D $ reaches a maximum value of the order $\sim 500$ eV when the shock arrives at the center (which occurs around time $t\approx 17500$ ps). Then, the discrepancy tends to disappear as one approaches the ignition threshold. This discrepancy has been observed experimentally\cite{ROS14}, and may be related to species separation effects inside the hot spot\cite{BEL14}. Indeed, the temporal evolution of the central densities $n_D$ and $n_T$ represented in Fig.\ref{fig:tite_implo}-right, for D and T species respectively, reveals that $n_D$ tends to rise slightly earlier than $n_T$ during the implosion process as the shock arrives in the central hot spot region. At later times, when collisional times decrease, the central species density discrepancy tends to vanish.




\begin{figure}[!h]
\centering
{\includegraphics[width=0.5\textwidth]{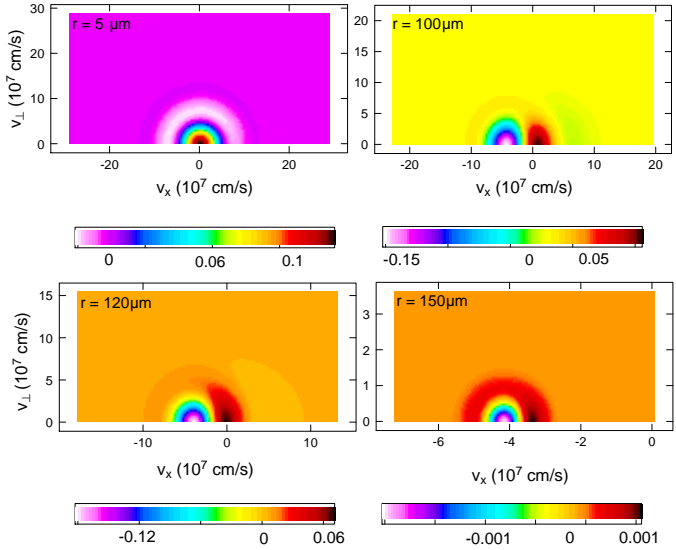}}
\caption{\label{fig:nlte_implo2} Non-LTE effects during the implosion: the kinetic modeling reveals a significant deviation from the Maxwellian behavior in the tail of the distribution during implosion at $t=17800$ ps. The map represents the normalized discrepancy $(\mathcal{M}_i(n_i,U_i,T_i)-f_i)/f_i^{max}$ between the local Maxwellian and the ion distribution function calculated by our kinetic code \textsc{Fuse}.}
\end{figure}

\begin{figure}[!h]
\centering
{\includegraphics[width=0.3\textwidth]{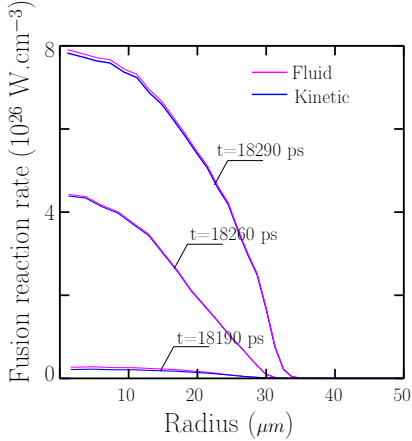}}
\caption{\label{fig:taurea_bin} Fusion rate spatial profiles calculated with respect to the non-Maxwellian distribution ion functions (blue) and to the equivalent local Maxwellian functions (magenta). }
\end{figure}

To study in more detail the deviation of  thermal ion distribution functions from the Maxwellian distribution function, one represents the normalized discrepancy $(\mathcal{M}_i(n_i,U_i,T_i)-f_i)(r,v_r,v_\bot,t)/f_i^{max}$ between the Maxwellian and the ion distribution function calculated in the kinetic model. The difference is normalized by the maximum value of the ion distribution function. We plot this quantity for $i=D$ in the thermal velocity grid, at time $t=17800$ ps, and at different locations of the fuel (see Fig.\ref{fig:nlte_implo2}). A deviation from the LTE behavior is visible in the center. This effect is related to the arrival of the shock wave that triggers ion-kinetic effects. This is also consistent with a higher ion heat flux observed in the kinetic calculation. Besides, a slight tail depletion effect is visible in the vicinity of the interface between the hot spot and the dense shell (see  Fig.\ref{fig:nlte_implo2}-bottom) . This is related to non-local transport effects, naturally taken into account in the kinetic modeling. Nonetheless, the amplitude of the deviation is rather small and may not have a significant impact on the local reactivity. Indeed, we compare in Fig.\ref{fig:taurea_bin} the kinetic fusion rate calculated  from the exact ion distribution functions, with the fusion rate estimated from the equivalent local Maxwellian functions. We do not see any significant effect on the fusion rate spatial profiles during the birth of the burn front.

Ion kinetic effects during implosion thus perturb ion thermal distribution functions, and have a clear impact on ion viscosity and ion heat flux, especially as the shock wave propagates through the hot spot and reflects from the center. This influences hydrodynamic profiles and the beginning of the combustion process. However, we do not find that the deviation of the distribution functions from the Maxwellian functions has a significant effect on fusion rate calculations.  

\subsection{Kinetic effects during combustion}

We now study the ion kinetic effects induced by  suprathermal $\alpha$-particles interacting with the thermal bulk during the combustion phase. We show that our full and self-consistent kinetic approach specially developed for fast fusion particles reveals significant effects on the birth and propagation of the burn front. 

\begin{figure}[!h]
\centering
{\includegraphics[width=0.5\textwidth]{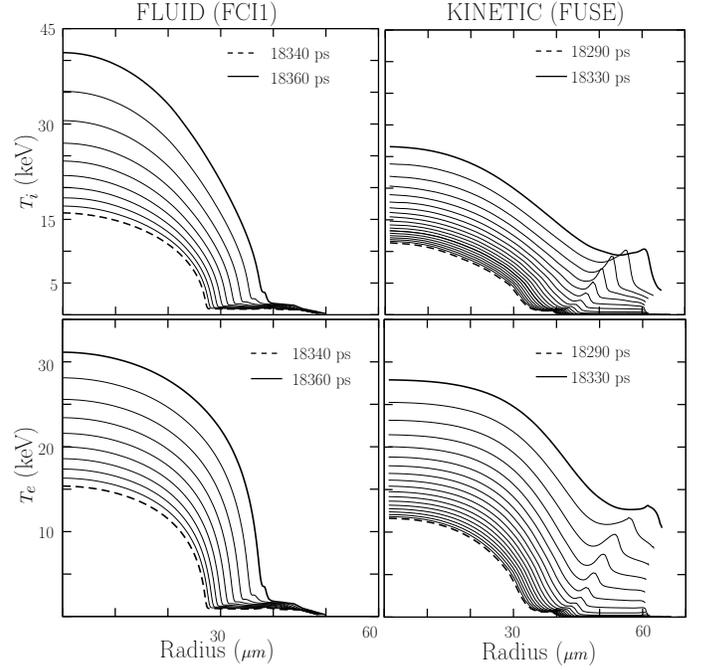}}
\caption{\label{fig:tite_burn1} Comparison of the fluid(\textsc{Fci1}) (left) and kinetic(\textsc{Fuse}) (right) electron-ion temperature spatial profiles during the beginning of the combustion, at times corresponding to the propagation of the flame through the dense fuel. }
\end{figure}

To study kinetic effects during the beginning of the combustion process, we start by plotting  the electron and ion temperature spatial profiles during the beginning of the combustion process when $\alpha$-particles start to play a significant role. Due to the loss of synchronism that occurs during the implosion phase, the observation times are different between the fluid and the kinetic calculation. The initial time corresponds to the maximum of the fuel areal density, whereas the final time is related to the maximum instantaneous power released by fusion reactions. The observation is thus performed in the vicinity of ignition, when the central hot temperature exceeds 10 keV and the fusion power is rising.  Comparisons are shown in Fig.~\ref{fig:tite_burn1}. 
In the kinetic calculation, a well defined \textit{pre-heating wave} tends to develop inside the dense fuel shell, while the central hot spot ion temperature remains significantly lower than in the fluid simulation.

\begin{figure}[!h]
\centering
{\includegraphics[width=0.5\textwidth]{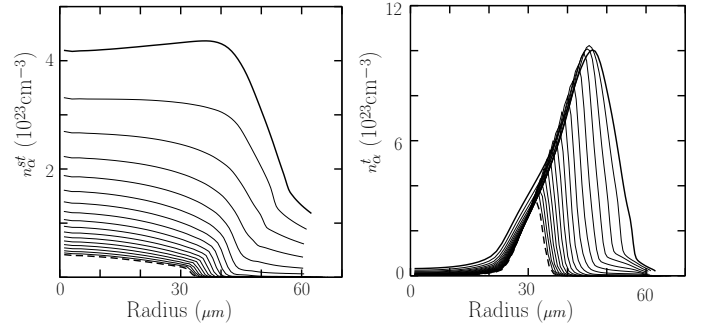}}
\caption{\label{fig:nih_nic} Two-component $\alpha$-particle density spatial profiles during the beginning of the combustion process in the kinetic simulation. One clearly sees that thermalized $\alpha$-particles accumulate in the region of the temperature precursor represented in Fig.~\ref{fig:tite_burn1}-right.  
Left: suprathermal $\alpha$-density profile; Right: thermal $\alpha$-density profile.
Dashed line: $t=18290$ ps; solid line: $t=18330$ ps. Times refer to the beginning of the implosion process.  
}
\end{figure}

To explain this new phenomenology, which is not present in the fluid modeling, we make use of our kinetic multi-scale approach and study the evolution of the \textit{suprathermal} and \textit{thermalized} $\alpha$ density spatial  profiles (referred to as $n_\alpha^{ST}$ and $n_\alpha^{T}$ respectively). The corresponding spatial profiles are represented in Fig.~\ref{fig:nih_nic} during the beginning of the combustion process.
The suprathermal density profiles shows that fast $\alpha$-particles are mainly created in the central hot spot, where the temperature is larger. Energetic particles subsequently deposit their energy in the surrounding cold shell. The region corresponding to the energy deposition associated to $\alpha$-particles is indicated by a sharp decreasing of the suprathermal density. Correspondingly, the bump  clearly visible on the thermal density profiles is related to the accumulation of thermalized particles. This occurs at a distance which corresponds to the collisional mean free path of suprathermal $\alpha$-particles. 
More quantitatively, a typical estimate of the $\alpha$-mean free path applicable to ICF conditions is given 
by \cite{FRA744}~:
$\rho_{hs} \lambda_\alpha \sim 0.1$ g.cm$^{-2}$. We can then apply this approximation to the conditions simulated in our kinetic modeling: for instance, one has at the time $t=18290$ ps, corresponding approximately to the maximum fuel areal density in the kinetic calculation: $\rho_{hs} \sim 30$ g.cm$^{-3}$ and $T_{hs}\sim$ 10 keV. Such conditions imply that $\lambda_\alpha \sim$ 30 $\mu$m. This is exactly what we observe on the thermal density profiles (see Fig.\ref{fig:nih_nic}-right) where thermalized $\alpha$-particles accumulate in a region located at a distance  $\sim 30$ $\mu$m.

Comparing ion temperature $T_i$ profiles (Fig.\ref{fig:tite_burn1}-top right) with the thermalized $ \alpha$-density $n_\alpha^T$ profiles (Fig.\ref{fig:nih_nic}-right) during the combustion in the kinetic modeling, one observes a clear correlation between the preheating wave that builds up ahead of the main temperature front and the spatial accumulation of thermalized $\alpha$-particles. Those particles that have slowed down on electrons then heat the thermal D,T ions located in the $\alpha$-particle accumulation region, close to the inner surface of the dense shell. This phenomenology is also linked to the well known Bragg's peak effect\cite{HON931}, such that slowed down $\alpha$-particles deposit their energy essentially on thermal ions, when they approach the thermal velocity domain. This effect is related to the  $\sim 1/v^2$ scaling in the $\alpha$-ion collisional drag term, see Eq.(\ref{eq:modele_1d}).

\begin{figure}[!h]
\centering
{\includegraphics[width=0.25\textwidth]{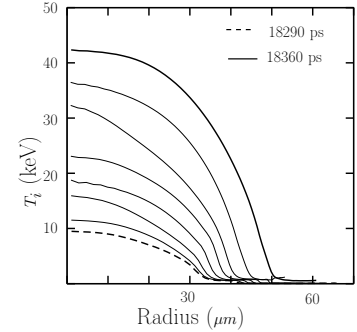}}
\caption{\label{fig:tisansadvec} Ion temperature spatial profiles obtained in the kinetic calculation where the $\alpha$-transport is artificially switched off.
}
\end{figure}

\begin{figure}[!h]
\centering
{\includegraphics[width=0.4\textwidth]{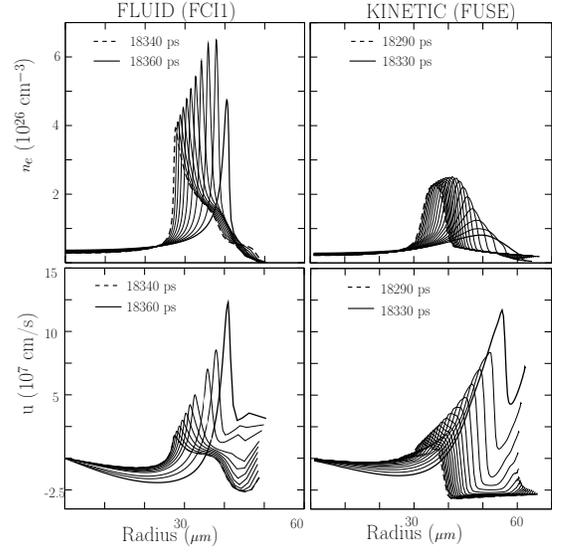}}
\caption{\label{fig:une_burn1} Comparison of the fluid (\textsc{Fci1}) (left) and kinetic(\textsc{Fuse}) (right) density-velocity spatial profiles during combustion, at times corresponding to the propagation of the flame through the dense fuel.}
\end{figure}

The pre-heating wave propagating through the dense fuel shell is thus linked to the kinetic enhancement of the suprathermal $\alpha$-particle transport.  Conversely, the fluid calculation, using a standard diffusion-like approach, tends to artificially trap energetic $\alpha$-particles inside the hot spot, thereby rising the temperature - and subsequently the reactivity - of the system. To support this interpretation and focus on the transport effect, we perform a kinetic calculation where the transport of energetic $\alpha$-particles is artificially switched-off. The corresponding ion temperature profiles are represented in Fig.\ref{fig:tisansadvec}. Not surprisingly, the spatial profiles appear to be more localized inside the hot spot, whose central region naturally reaches higher temperatures. One thus recovers spatial profiles that are close to the fluid-diffusion scheme model (see Fig.\ref{fig:tite_burn1}-top left).  Besides, kinetic effects during the implosion phase are such that the heating of the hot spot happens faster. Consequently, the kinetic combustion occurs in a less dense and larger hot spot, as it is shown on the density profiles in Fig.\ref{fig:une_burn1}, top panel. Since the regions corresponding to the suprathermal $\alpha$-particle energy deposition  are located farther outside the hot spot in the kinetic modeling,  the dense fuel is characterized by higher mean velocities in the kinetic simulation (see Fig. \ref{fig:une_burn1}-bottom panel).

\begin{figure}[!h]
\centering
{\includegraphics[width=0.5\textwidth]{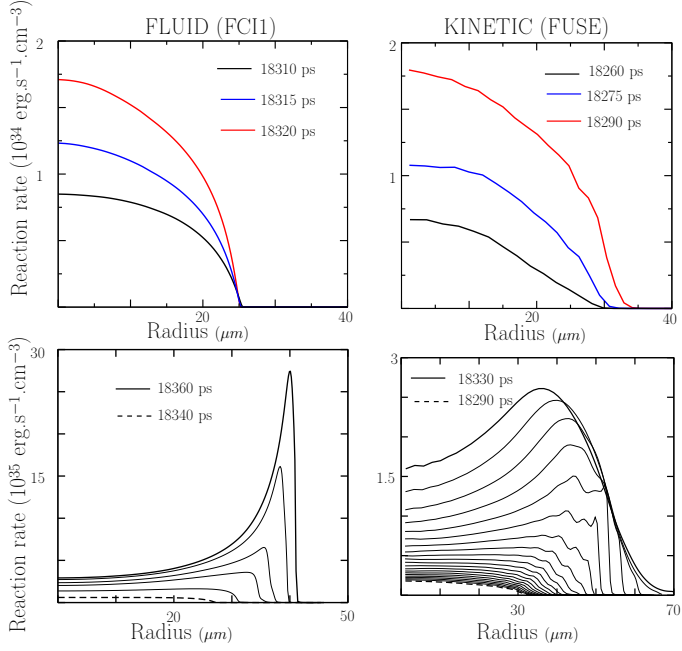}}
\caption{\label{fig:burn_front} Comparison of the burn front structure during its birth and its propagation through the dense fuel shell between the fluid (left) and kinetic modeling (right).}
\end{figure}

In order to illustrate  the consistency of our kinetic results and to explain in more detail the combustion process, it is instructive to study the development of the burn front in the fluid model, based on a diffusion-like scheme, and in our two-velocity-scale kinetic approach. The reaction rate spatial profiles are represented  at different times corresponding to the beginning of the combustion in both calculations in Fig.\ref{fig:burn_front}-top.  The flame appears much more localized inside the hot spot in the fluid model, whereas  the kinetic reaction rate spatial profiles reach a region located farther inside the dense fuel shell. During the burn front propagation, the difference increases (Fig.\ref{fig:burn_front}-bottom): the kinetic flame front is clearly less sharp and broader than the fluid one, whose spatial profiles follow the density profiles given in Fig.\ref{fig:une_burn1}-top left panel.

\begin{figure}[!h]
\centering
{\includegraphics[width=0.4\textwidth]{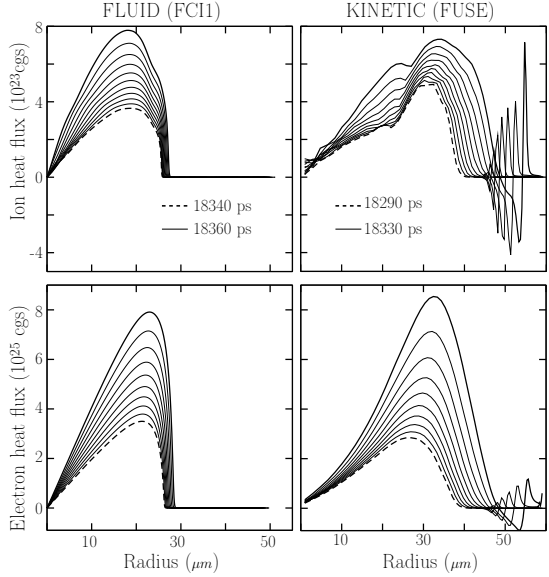}}
\caption{\label{fig:qiqe_burn} Comparison of the ion (top) and electron (bottom) heat flux during the flame propagation between the fluid (left) and kinetic (right) simulations.}
\end{figure}

Moreover, the $\alpha$-particle transport enhancement observed in the kinetic modeling significantly modifies heat flux profiles during the flame propagation (Fig.~\ref{fig:qiqe_burn}). In particular, the ion heat flux is characterized by a sharp precursor structure propagating ahead of the main front. This phenomenology is consistent with the pre-heating wave observed in the ion temperature profiles in the kinetic calculation. Besides, the impact of the enhanced $\alpha$-particle transport  is also clearly visible on electron heat flux profiles during the first part of the combustion process (see Fig.~\ref{fig:qiqe_burn}-bottom), corresponding to the propagation of the burn front through the dense fuel shell.

\subsection{The combustion process at a kinetic level}

\begin{figure}[!h]
\centering
{\includegraphics[width=0.45\textwidth]{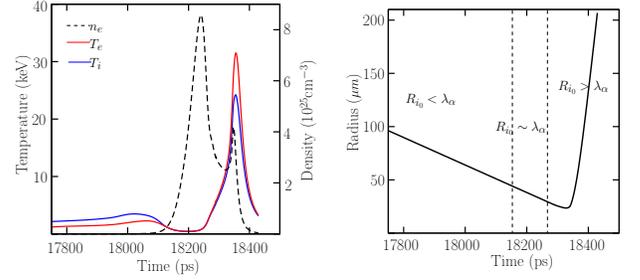}}
\caption{\label{fig:mesh_ix15} Thermodynamic conditions (left) and radius (right) of the test cell chosen inside the hot spot to study the local combustion process in the kinetic calculation.
}
\end{figure}

\begin{figure}[!h]
{\hspace{-0.3cm}\includegraphics[width=0.5\textwidth]{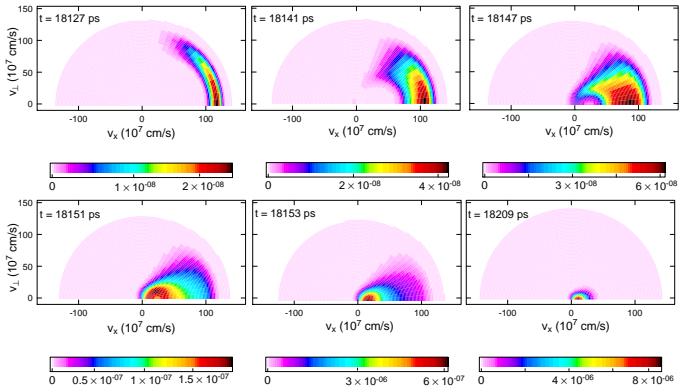}}
\caption{\label{fig:fdh_seq1} Slowing down of suprathermal particles observed in the considered test cell.
}
\end{figure}

In this section, we focus on the coupling between the suprathermal and thermal components associated to $\alpha$-particles during the combustion process. This gives an illustration of the relevancy of the two-velocity-scale approach. We consider a given spatial test cell $i_0$ chosen inside the hot spot in the kinetic calculation. The radius of the considered cell $r_{i_0}(t)$  is represented as a function of time in Fig.~\ref{fig:mesh_ix15}-right. The suprathermal distribution function of $\alpha$-particles   $f_\alpha^{ST}(r_{i_0}(t),v,\theta,t)$ observed in the considered mesh is given in Fig.~\ref{fig:fdh_seq1} at different observation times.

Let us discuss the temporal evolution of the suprathermal component represented  in Fig.~\ref{fig:fdh_seq1} during the end of the implosion process.
The suprathermal component is rather anisotropic. It is highly peaked towards positive velocities $v_r >0$. This can be explained by the inhomogeneous fusion reaction source term, which is peaked towards the hot center of the capsule. Since the considered cell is located outside of the central emissive zone, we see $\alpha$-particles passing from the center (which is on the left of the considered cell) to the outside (which is on the right of the considered mesh). In other words, $\alpha$-particles deposit their energy at a distance which is greater than the radius of the considered cell.

Then, during the implosion process, there exists a time when the radius of the considered test cell becomes close to one $\alpha$-mean free path (see Fig.\ref{fig:mesh_ix15}-right). At this time, $\alpha$-particles deposit their energy in the considered cell: the $\alpha$ distribution function then slows down towards the thermal velocity domain. As it slows down,  the distribution function tends to spread over a wider domain in the polar angle direction. This spreading is the consequence of the diffusion part of the Fokker-Planck operator, leading to a mainly transverse slowing down current that intensifies as the $\alpha$-particles get closer to the thermal velocity region.

\begin{figure}[!h]
\centering
{\includegraphics[width=0.45\textwidth]{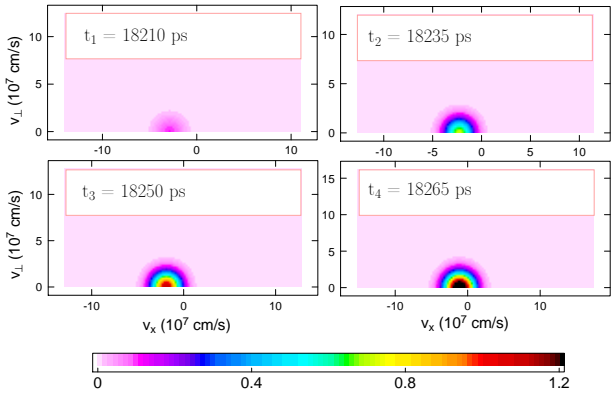}}
\caption{\label{fig:fdc_feeding} Feeding of the thermal component by the suprathermal one. This represents the $\alpha$-thermal component on the thermal grid, observed in the considered spatial test cell.
}
\end{figure}

When slowed down suprathermal particles reach the thermal velocity domain, 
they feed a thermal component which is represented  on the thermal velocity grid in Fig.~\ref{fig:fdc_feeding}. 
The $\alpha$ thermal component then interacts with the other thermal species to complete its relaxation process on the thermal velocity scale.

\begin{figure}[!h]
\centering
{\includegraphics[width=0.45\textwidth]{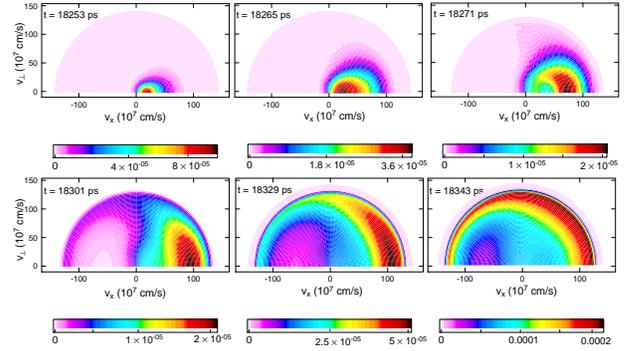}}
\caption{\label{fig:fdh_combu}Temporal evolution of the $\alpha$-suprathermal component as the combustion takes place in the considered test cell.
}
\end{figure}
Finally, during the beginning of the combustion process, suprathermal $\alpha$-particle production intensifies, the local temperature increases and the local density decreases (as it is shown in Fig.\ref{fig:mesh_ix15}-left), so that the considered cell tends to become transparent to $\alpha$-particles once again, and an anisotropic suprathermal component with significantly higher values builds up (see Fig.\ref{fig:fdh_combu}). 

\section{Summary and perspectives}

In this article, we have presented a tractable full ion kinetic approach, applicable to real ICF configurations during the implosion, ignition and burn stages. The model is based on a two-scale decomposition of the suprathermal distribution function. It is then feasible to tackle fusion products accurately at a ion-kinetic level and in a self consistent manner with the thermal bulk. The transport, acceleration and collisional effects related to $\alpha$-particles are then described with a better precision than in usual fluid codes where a diffusion scheme is employed. We recall that the diffusion approximation may not be applicable to energetic particles since their mean free path is comparable to the hot spot radius.

We have then presented full ion kinetic simulations of a typical ICF target and compared the results with the fluid model.
Ion kinetic effects start during implosion, especially as the shock wave arrives at the hot spot center. It has been shown that  ion distribution functions tend to deviate from  Maxwellian functions. The ion heat flux is then modified and the heating of the hot spot tends to build up faster in the kinetic modeling.  

Consequently, the combustion tends to start earlier in the kinetic model, in a less dense and larger fuel. During the combustion process, ion-kinetic effects on suprathermal particles have a significant impact on the burn front structure. The non-local transport of fast particles is enhanced in the kinetic model, so that suprathermal $\alpha$-particles deposit their energy and momentum mainly \textit{out of the hot spot}, in a region corresponding to the inner surface of the dense shell, located at a distance corresponding to one $\alpha$-mean free path (which is close to the hot spot radius). Conversely, energetic particles happen to be trapped inside the hot spot in the fluid calculation where the diffusion approximation may tend to artificially localize $\alpha$-energy deposition. 

The perturbation of burn front structures due to non-local transport effects subsequently modify hydrodynamic profiles during the flame propagation. The full kinetic approach reveals the existence of a \textit{pre-heating wave}, that propagates ahead of the main burn front, clearly visible on ion temperature spatial profiles. We have shown that the precursor structure is a consequence  of the Bragg's peak effect related to the $\alpha$-energy deposition on thermal ions. At a global level, ion-kinetic effects significantly reduce the central temperature, pressure and fusion yield, by $\sim 50\%$.

This new effect may contribute to account for the difficulties encountered  in present day ICF target designs to reach ignition. However, this does not explain why DT capsules do not ignite, because our kinetic simulations lead to an ignition, although the subsequent combustion is less efficient. There may be other causes (instabilities, DT/pusher mix, asymmetry, capsule/hohlraum interaction ...) possibly coupled with the non-local effects associated to fusion products, that may be invoked to properly explain the non-ignition. Nonetheless, the phenomenology presented here based on a full kinetic approach may have significant consequences on future ICF target designs, that should thoroughly take into account the non-local feature of the momentum and energy deposition associated to suprathermal particles.    

Besides, the non-local $\alpha$-energy deposition may have an influence on the DT fuel /CH pusher mix. Indeed, the pre-heating wave supported by $\alpha$-particles may tend to interact with possible fuel/pusher mix structures (such as CH spikes mixing with the DT shell, in the case of Rayleigh-Taylor instability) thereby modifying the development of the instability. To analyze in more detail the interaction between the CH pusher and the DT fuel, the modeling of the CH pusher should be consistently taken into account at the kinetic level. That may require the development of a hybrid code, treating some of the pusher species at a fluid level (for instance the carbon species, which is highly collisional) as well as modeling D,T,$\alpha$ and H species at an ion-kinetic level (possibly with two-velocity components).

Finally, the two-component formalism devised for $\alpha$-particles could be naturally extended to add the effect of Boltzmann-type large angle scattering, that would feed a suprathermal component for the D-T ions.  Neutron momentum and energy deposition may also be modeled in a similar way.
Those extensions are left for future work.


\end{document}